\documentclass[11pt]{article}
\usepackage{algorithmicx, algorithm, algpseudocode,amsmath, amsthm, amssymb, authblk, bm,caption, datatool, enumerate, float, glossaries, graphicx, graphics, hyperref, index,  listings, multirow, nomencl, natbib,ltablex, longtable,  rotating, scalerel,  setspace, subfig,tabularx, titlesec,url, verbatim, 
}
\usepackage[table,xcdraw]{xcolor}
\usepackage[a4paper,inner=2cm,outer=2cm,top=2.5cm,bottom=2.5cm,pdftex]{geometry}

\usepackage[english]{babel}
\hypersetup{
    colorlinks,
    linkcolor={blue!80!black},
    citecolor={blue!80!black},
    urlcolor={blue!80!black}
}
\graphicspath{ {figures/} }
\algnewcommand{\LineComment}[1]{\State \(\triangleright\) #1}

\DeclareMathOperator{\Var}{\mathbb{V}ar}
 \DeclareMathOperator*{\argmin}{arg\,min}

\title{Nonmyopic and pseudo-nonmyopic approaches to optimal sequential design in the presence of covariates}

\author[1]{Mia Sato Tackney}
\author[1]{David C. Woods}
\author[3]{Ilya Shpitser}

\affil[1]{London School of Hygiene and Tropical Medicine}
\affil[2]{University of Southampton}
\affil[3]{Johns Hopkins University}

\begin{document}
\maketitle

\section*{Abstract}
In sequential experiments, subjects become available for the study over a period of time, and covariates are often measured at the time of arrival. We consider the setting where the sample size is fixed but covariate values are unknown until subjects enrol. Given a model for the outcome, a sequential optimal design approach can be used to allocate treatments to minimize the variance of the treatment effect. We extend  existing optimal design methodology so it can be used within a nonmyopic framework, where treatment allocation for the current subject depends not only on the treatments and covariates of the subjects already enrolled in the study, but also the impact of possible future treatment assignments. The nonmyopic approach is computationally expensive as it requires recursive formulae. We propose a pseudo-nonmyopic approach which has a similar aim to the nonmyopic approach, but does not involve recursion and instead relies on simulations of future possible decisions. Our simulation studies show that the myopic approach is the most efficient for the logistic model case with a single binary covariate and binary treatment. \\

Keywords: design of experiments, optimal design, dynamic programming, sequential design, coordinate exchange

\section{Introduction}
How treatments should be allocated in sequential experiments in the presence of covariates is a highly debated topic, particularly within the clinical trials community \citep{Senn2013, Rosenberger2008}. We consider experiments where subjects become available sequentially, covariates are measured at the time of arrival, and treatment is assigned soon after. We assume that a response is measured before the next subject arrives, and we assume a fixed sample size. At any point in the experiment, the covariate values for the subjects yet to enrol in the experiment are unknown. Such a set-up is often characteristic of large Phase III trials, but is also common in experiments in the social sciences, such as political psychology lab experiments \citep{Moore2013}. Covariates should be included in the analysis; omitting them results in bias \citep{Senn2013}, and further, from an optimal design point of view, the allocation of treatment should be done in a way that maintains as equal replication as possible of treatment within covariate groups, which improves precision of the parameter estimates \citep{Atkinson1982}. \\ 

Minimization is an approach aimed to keep the numbers of treatments approximately equal for each group of subjects who have the same combination of covariate values and is now used extensively in clinical trials \citep{Pocock1975, Taves1974}. It has received some criticism for being based on measures of imbalance of covariates which are not theoretically grounded \citep{Senn2010} and methods based on minimizing the variance of the parameter estimators in statistical models have been suggested instead, originally by \citet{Atkinson1982}. Atkinson's optimal design approach aims to minimize the variance of the estimator of the treatment effect for a linear model which describes the relationship between the treatments, covariates and response. The $D_A$-optimal objective function is used to make decisions for treatment allocation. We generalize Atkinson's approach for the logistic model case, which can be applied to any information matrix-based optimality criterion in Section \ref{myopic}. \\\

The sequential optimal design approach is myopic in the sense that decisions are made using information about the past subjects' covariates, treatments and response and the current subject's covariates. The decision about the current subject is made assuming that the experiment will terminate after its response is recorded, ignoring the fact that there are further subjects which will enter the trial, and the estimates of interest are based on data from the entire experiment. Nonmyopic approaches are able to consider the potential impact of the current treatment decision on future possible decisions \citep{Huan2016} in terms of efficiency of the estimators. In this paper, we assess whether there is a benefit, in terms of efficiency of the estimators, in taking into account the impact of future possible decisions.  This relies on the method of dynamic programming to compute the expected value of the objective function, where the expectation is taken over unknown quantities of future subjects \cite[p. 323]{Bradley1977}. Most applications of nonmyopic approaches in clinical trials aim to maximize some measure of benefit of the treatment to the subject. The Gittens index is an example of such a nonmyopic approach \citep{Gittens1979, Smith2018, Williamson2017, Villar2018}. Nonmyopic approaches for a clinical trials based problem involving covariates where the objective is related to the estimation of parameters have not been explored explicitly in the literature. We address this in Section \ref{nonmyopic} and compare the myopic and nonmyopic approaches in a simulation study. \\
 
The nonmyopic approach is computationally expensive which limits its use in practical settings. We propose the pseudo-nonmyopic approach in Section \ref{pseudononmyopic}, which has a similar aim to the nonmyopic approach but does not require recursive formulae. We compare how it fares against the myopic approach in a simulation.  We discuss our findings and potential extensions of our work in Section \ref{discussion}.

\section{Myopic Sequential Design} 
\label{myopic}
\subsection{Optimal Design}
Suppose there are $n$ subjects in total in an experiment, which is fixed from the start. For $i \in \left\{1, ..., n\right\}$, we observe the values of the $s$ covariates associated with unit $i$

\begin{equation}
 \bm{z}_i= \begin{pmatrix} z_{i, 1}, ..., z_{i, s} \end{pmatrix}^T, 
\end{equation} 

and we select a treatment $t_i$ from a set of possible treatments $\mathcal{T}$. We observe the response $y_i$, which we assume is binary and zero is the desirable response. We define the following:

\begin{equation*}
\bm{Z}_i = \begin{pmatrix}
\bm{z}_1^T \\
 \bm{z}_2^T \\
  \vdots \\
 \bm{z}_i^T
\end{pmatrix} ,
\end{equation*}

\begin{equation*}
\bm{t}_i = \begin{pmatrix}
t_1, t_2, ... , t_i
\end{pmatrix}^T ,
\end{equation*}

\begin{equation*}
\bm{y}_i = \begin{pmatrix}
y_1, y_2, ... , y_i
\end{pmatrix}^T ,
\end{equation*}
to be the $i \times s$ matrix of covariate values, the $i$-vector of treatments and $i$-vector of responses, respectively, for subjects 1 up to $i$.\\

Given that $y_i$ has distribution 

\begin{equation}
\label{dist}
y_i \sim \mbox{Bernoulli}(\pi_i), 
\end{equation}

we assume a logistic regression for the response, where the probability $\pi_i=\mathbb{P} \left( y_i=1 \right)$ is given by

\begin{equation}
\pi_i = \frac{ \exp{\eta_i}}{1+\exp{\eta_i}}, 
\end{equation}

where $\eta_i$ is the linear predictor. We assume $\eta_i$ is a linear combination of the intercept, main effects for the covariates and treatment, and potentially interaction terms. We denote the number of terms in the linear predictor by $q$ and assume it takes the following form:

\begin{equation}
\label{predictor}
\eta_j =  \bm{x}_j \bm{\beta},
\end{equation}

where $\bm{x}_j$ is the $j$th row of the $i \times q$ design matrix $\bm{X}_i$ , for $j \in \left\{1, ..., i \right\}$, and we denote by $\bm{\beta}$ the associated $q$-vector of parameter values. We can write the information matrix as $\bm{I} = \bm{X}_i^T \bm{W}_i \bm{X}_i$, where $\bm{W}_i$ is a diagonal matrix with $(j, j)$th entry given by $\hat{\pi}_j(1-\hat{\pi}_j)$.  There is a close link between the information matrix and the variance of the parameter values; the maximum likelihood estimator of ${\bm{\beta}}$ has asymptotic variance-covariance matrix given by the inverse of the information matrix: 

\begin{equation*}
\Var(\bm{\beta}) =\left( \bm{X}_i^\top \bm{W}_i \bm{X}_i \right) ^{-1}.
\end{equation*} 

In optimal design theory, decisions about treatments are made to minimize some function of $\left( \bm{X}_i^\top \bm{W}_i \bm{X}_i \right) ^{-1}$. A $D$-optimal design minimizes the determinant of the inverse of the information matrix, or equivalently, it minimizes the volume of the confidence ellipsoid of $\bm{\beta}$ \citep[p. 53]{Atkinson2007}. The $D$-optimal objective function, assessing the choice of treatments of the subjects enrolled in the study $t_1, ..., t_i$, is given by:

\begin{equation}
\label{D_logit}
 \Psi_{D}(\bm{X}_i, \bm{\beta})=\left| \left( \bm{X}_i^T \bm{W}_i \bm{X}_i\right)  ^ {-1} \right|, 
\end{equation}

where $\left| \cdot \right|$ denotes the determinant. One may have interest only in a subset of the parameters, or in some linear combination of them. Supposing that interest lies in $m$ linear combinations of $\bm{\beta}$, the quantity of interest can be expressed as $\bm{A}^T \bm{\beta}$, where $\bm{A}$ is a $q \times m$ matrix with $m < q $ \citep[p.137]{Atkinson2007}. The asymptotic variance-covariance matrix for $\bm{A}^T \bm{\beta}$ is given by: 

\begin{equation}
 \Var \left( \bm{A}^T \bm{\beta} \right) =  \bm{A}^T \left( \bm{X}_i^T \bm{W}_i \bm{X}_i\right)  ^ {-1} \bm{A} .
\end{equation}

In this case, the $D_A$-optimality criterion is more appropriate:

\begin{equation}
\label{D_A_logit}
 \Psi_{D_A}(\bm{X}_i, \bm{\beta})= \left| \bm{A}^T \left( \bm{X}_i^T \bm{W}_i \bm{X}_i\right)  ^ {-1} \bm{A} \right|.
\end{equation}

In our case, we are interested in estimating the treatment effect as precisely
as possible, $\bm{A}$ is a $q$-vector with entry one corresponding to the treatment effect and zeros otherwise. If we wish to evaluate the decision for the $i$th treatment given the covariates of subjects 1 up to $i$, $\bm{Z}_{i}$, the treatments of the previous subjects, $\bm{t}_{i-1}$, and the responses of previous subjects, $\bm{y}_{i-1}$, we can denote the value of a generic objective function evaluated when treatment $t_i$ is assigned to subject $i$ as

\begin{equation}
\label{seq_obj}
 {\Psi} ( t_{i} \mid  \bm{Z}_{{i}}, \bm{t}_{i-1},  \bm{y}_{i-1}),
\end{equation}

where $\Psi$ could be the $D$- or $D_A$-, or some other information matrix based objective function. In a non-sequential setting, an optimal allocation of treatments $\bm{t}_i$ for a design with $i$ subjects $\bm{X}_i$ can be constructed using the exchange algorithm. See, for example, \citet[p.36]{Goos2011} or \citet[p.172]{Atkinson2007}.\\

In a sequential setting, extending the work of \cite{Atkinson1999} for binary treatments, we assign treatment $t \in \mathcal{T}$ to subject $i$ by the probability given by 

\begin{equation}
\label{logit_tmt1}
\frac{ {\Psi} ( t_{i}=t \mid \bm{Z}_{{i}}, \bm{t}_{i-1},  \bm{y}_{i-1}) ^ {-1}} { \sum_{t \in \mathcal{T}} {\Psi} ( t_{i}=t \mid \bm{Z}_{{i}}, \bm{t}_{i-1},  \bm{y}_{i-1}) ^ {-1} }.
\end{equation}

This is a biased-coin type approach to treatment allocation, where the optimal decision according to the objective function $\Psi$ is likely to be selected, but there is random variation to avoid any suspicion of selecting bias \citep{Atkinson1982}. \\

The evaluation of the objective function in \eqref{seq_obj} can be problematic in the context of logistic regression for two reasons. Firstly, because we have binary treatments and potentially binary covariates, separation is more likely to occur, where a linear combination of covariates perfectly predicts the response (\citealp{Firth1993}, \citealp{Gelman2008}). Separation can result in the likelihood function becoming monotonic and maximum likelihood estimates of the regression coefficients tending to plus or minus infinity \citep{Rainey2016}. Common approaches of dealing with separation include penalizing maximum likelihood estimates to reduce bias, and introducing a prior distribution for the regression coefficients to shrink parameter estimates, particularly large ones, to zero. Jeffry's prior is a common choice of prior, and \citet{Gelman2008} recommended independent Cauchy distributions, where the probability distribution function given the location parameter $x_0$ and scale parameter $\gamma$ is given by: 

\begin{equation}
f\left(x \mid x_0 \gamma, \right) =\frac{1}{\pi \gamma \left( 1+\left(\frac{x-x_0}{\gamma}\right)^2 \right)    }
\end{equation} 

where the intercept has $x_0= 0, \gamma= 10$, and the slope coefficients have $x_0= 0, \gamma= 2$. We use this recommendation by \citet{Gelman2008}.\\

Secondly, the objective function in \eqref{seq_obj} depends on values of the model parameters. Therefore, estimates of parameters are needed in order to design the experiment aimed to estimate these parameters in the first place \citep{Atkinson2015}. We overcome this by beginning with an initial design where we use the exchange algorithm to allocate treatments for an initial $n_0$ units, under the assumption that $\bm{\beta}$ is a vector of zeros. Responses are then obtained or generated for the first $n_0$ subjects, and the model is fit to obtain the first estimate of the model parameters. Algorithm \ref{Algorithm1} in the Appendix outlines the steps in constructing a sequential optimal design.

\section{Nonmyopic Approach} 
\label{nonmyopic}

Having a nonmyopic approach to the treatment allocation problem means that the optimization involves multiple stages. Not only is it important to consider the impact of the decision at the time of subject $i$, but we consider future subjects, possibly up to subject $n$. The number of future subjects considered is called the horizon, denoted $N$.  The state at stage $i$ comprises the information that is known at that stage, which in our example includes the values of the covariates of subjects 1 up to $i$, as well as the treatments and responses of subjects 1 up to $i-1$. The decision about the treatment $t_i$ at stage $i$ is made based on the state $S_{i-1} = (\bm{Z}_{i},\bm{t}_{i-1}, \bm{y}_{i-1}).$ Based on that decision, there is a transition function $f_i$ that outputs the state of the next stage, $S_{i}=f_i(S_{i-1}, t_i)$. In our case, this transition function is represented by a logistic model linking the responses to the treatments and covariates. There is a need to balance two conflicting aims in the decision making: 

\begin{enumerate}
\item The aim to exploit and to choose the treatment that most precisely estimates $\bm{\beta}$ in the current state.

\item The aim to explore and to choose treatment which may not be optimal given the current state, but may lead to gain of information which could lead to more precise estimators in later states.  
\end{enumerate}

Dynamic programming is an approach for solving multistage optimization problems (see, for example, \citealp{Powell2009}). The overall problem is broken into different stages, which often correspond to points in time, and each stage of the problem can be optimized conditionally on past states. The key idea is that the overall sequence of decisions for treatment selection will be optimal for the entire experiment \cite[p. 320]{Bradley1977}.  The optimal design can be obtained by forward or backward induction. We focus on backward induction since it is the approach that is usually most appropriate in problems involving uncertainties \citep[p. 328]{Bradley1977}. In backward induction, we start by finding the optimal decision at the end of the sequence of decisions, taking into account all possible treatments and covariates that may have been observed up until that point. Then, one can work backwards and obtain the optimal design taking expectations of unknown quantities \citep[p.330]{Bradley1977}. See \cite{Huan2016} for a recent overview of approximate dynamic programming in the context of Bayesian experimental design. Dynamic programming has been used in some clinical trials applications where one wishes to balance the aim of estimating the parameters (exploration) with the aim of giving subjects the best possible treatment or obtaining maximum total revenue (exploitation). See, for example,  \cite{Cheng2007}, \citet{Ondra2019}, \cite{Muller2007} or \citet{Bartroff2010}. \\

We now describe the nonmyopic approach for the binary response. To keep notation simple, we assume that we have a single binary covariate and a single binary treatment, and we do not consider interactions. We begin by constructing an initial design $\bm{X}_{n_0}$ with $n_0$ subjects  using the exchange algorithm. We assume $\bm{\beta}=\bm{0}$ as an initial guess for evaluating the objective function in the construction of $\bm{X}_{n_0}$. We then obtain responses for the first $n_0$ subjects, $\bm{y}_{n_0}$, and fit the model to obtain the initial maximum likelihood estimates of the model parameters, $\hat{\bm{\beta}}_0$. \\

Now, supposing that we have a design for $i-1$ subjects, and we have obtained parameter estimates $\hat{\bm{\beta}}_{i-1}$ as a result of that design. We observe covariate value $z_{i}$ for the $i$th subject and wish to evaluate the impact of assigning treatment $t_i$ on decisions on future possible subjects. For example, for horizon $N=1$, we consider the expected value of the objective function after $i+1$ subjects. Suppose treatment $t_i$ is assigned to subject $i$. Since $\Psi$ depends on $y_i$, we need to consider the two possible responses that $y_i$ may take, and then consider the possible values that ${z}_{i+1}$ can take. For a given covariate value ${z}_{i+1}$ for subject $i+1$, we denote by $t^*_{i+1}({z}_{i+1}, t_{i}, y_i \mid  \bm{z}_i,  \bm{t}_{i-1},   \bm{y}_{i-1})$ the optimal choice of treatment for subject $i+1$ given ${z}_{i+1}$ and $t_i$:

\begin{equation}
t^*_{i+1}({z}_{i+1}, t_{i}, y_i \mid  \bm{z}_i,  \bm{t}_{i-1},   \bm{y}_{i-1})  =\displaystyle  \argmin _{t_{i+1}} \Psi (t_{i+1} \mid \bm{z}_{i+1}, \bm{t}_{i}, \bm{y}_{i}).
\end{equation} 

From here on, we suppress the conditioning and write $t^*_{i+1}({z}_{i+1}, t_{i},  y_i)$ for simplicity. Now, we take the expectation of the objective function over two possible responses which may be obtained to find an expected value of the objective function over the unknown response:

\begin{align*}
\mathbb{E}_{{y}_{i}} \Psi (t_{i+1} \mid  \bm{z}_{i+1},   \bm{t}_{i},  \bm{y}_{i})& =  \mathbb{P} ({y}_{i}=0 \mid  \bm{z}_i,  \bm{t}_i,  \bm{y}_{i-1})  \Psi (t_{i+1} \mid  \bm{z}_{i+1},  \bm{t}_{i},  \bm{y}_{i-1},y ) \\ 
&+ \mathbb{P} ({y}_{i}=1 \mid  \bm{z}_i,  \bm{t}_i,  \bm{y}_{i-1})  \Psi (t_{i+1} \mid  \bm{z}_{i+1},  \bm{t}_{i},  \bm{y}_{i-1},y ) , 
\end{align*}

where $y_{i} \sim$ Bernoulli$(\pi_{i})$ with $\pi_{i}$ given by: 

\begin{equation}
\pi_{i} = \frac{\exp \left(\bm{x}_i \hat{\bm{\beta}}_{i-1} \right) }{1+\exp\left( \bm{x}_i \hat{\bm{\beta}}_{i-1}  \right)},
\end{equation}

where $\bm{x}_i = \begin{pmatrix} 1,  z_i,  t_i \end{pmatrix}$ is the $i$th row of the design matrix. Now, we consider the possible covariate values that we may observe for the next subject. We denote by $ \mathbb{P} ({z}_i = {z}) $ the probability that the $i$th subject has covariate value ${z}$. In some cases, the distribution of the covariates may be known; if not, the distribution can be estimated by the empirical distribution of the covariates of the first $i$ subjects.  We denote by $\Psi_1(t_i\mid  \bm{z}_i,  \bm{t}_{i-1},  \bm{y}_{i-1} )$ the expected value of the objective function when treatment $t_i$ is assigned to subject $i$, taking into account the impact of the decision on one further decision in the future. We obtain an expectation over the possible covariate combinations of the optimality criterion: 

\begin{align}
\Psi_1(t_i\mid  \bm{z}_i,  \bm{t}_{i-1},  \bm{y}_{i-1}) &= \mathbb{E}_{\bm{z}_{i+1}} \mathbb{E}_{{y}_{i}} \Psi (t^*_{i+1}(\bm{z}_{i+1}, t_{i}, y_i) \mid  \bm{z}_{i+1},   \bm{t}_{i},  \bm{y}_{i}) \\
&= \displaystyle \sum_{{z} } \mathbb{P} ({z}_{i+1} = {z}) \mathbb{E}_{{y}_{i}}  \Psi (t^*_{i+1}({z}_{i+1}, t_{i}, y_i ) \mid  \bm{z}_{i},  {z}_{i+1}, \bm{t}_{i}, \bm{y}_{i}  ) .
\end{align}

For a horizon greater than 1, we can use the following recursive relationship to find the optimal treatment for subject $i$. The expected value of the objective function after $i +N$ subjects, when treatment $t_i $ has been assigned, is given as follows: 

For $N > 0$:

\begin{align}
\label{logisfuture}
\begin{split}
\Psi_N(t_i &\mid  \bm{z}_i,  \bm{t}_{i-1},  \bm{y}_{i-1}) = \mathbb{E}_{\bm{z}_{i+1}} \mathbb{E}_{{y}_{i}} \Psi_{N-1}(t^*_{i+1}(\bm{z}_{i+1}, t_{i}, y_i) \mid  \bm{z}_{i+1},   \bm{t}_{i},  \bm{y}_{i}) \\
&= \displaystyle \sum_{{z} } \mathbb{P} ({z}_{i+1}= {z}) \mathbb{E}_{{y}_{i}}  \Psi_{N-1} (t^*_{i+1}(\bm{z}_{i+1}, t_{i}, \bm{y}_i ) \mid  \bm{z}_{i}, {z}_{i+1}, \bm{t}_{i},\bm{y}_{i}  ) ,
\end{split}
\end{align}

and for $N=0$, we have
\begin{align}
\Psi_0(t_i \mid  \bm{z}_i,  \bm{t}_{i-1},  \bm{y}_{i-1}) =\Psi(t_i \mid  \bm{z}_i,  \bm{t}_{i-1},  \bm{y}_{i-1}) ,
\end{align}

which is simply the myopic loss after $i$ subjects. We note that the nonmyopic approach for the logistic model case is considerably more computationally intensive than the myopic approach.

\subsection{Simulations}
\label{nonmy_simulations}
Our simulation compares $D_A$-optimal designs that are constructed sequentially using myopic and nonmyopic methods. Further, we compare the nonmyopic approach where we assume the true distribution for the covariates, and the nonmyopic approach where we use the empirical distribution of the covariates obtained by finding the proportion of observed subjects with each covariate value. \\

Since the information matrix and the objective function depend on values of the model parameters in the logistic model case, estimates of parameters are needed in order to design the experiment aimed to estimate these parameters in the first place. We begin with an initial design where we use the exchange algorithm to allocate treatments to 10 units, under the assumption that $\bm{\beta}$ is a vector of zeros. In order to reduce sources of variability in our simulations, we make sure that the same initial design is used for the myopic and non-myopic cases.  Another source of variability is the generation of the responses which are needed to obtain the estimates of the model parameters and subsequently to evaluate the design under the objective function. When comparing the myopic and non-myopic designs, we generate the responses in the following way: 

\begin{enumerate}
\item Generate a deviate $u_i$ from the Unif$(0,1)$ distribution.
\item Set 
\begin{equation}
\label{ueq}
 y_i = \begin{cases}1 & \mbox{if } u_i \geq \pi_i \\ 0 & \mbox{if } u_i < \pi_i \end{cases}.
\end{equation} 
\end{enumerate}

The deviates $u_i$ kept the same for the myopic and non-myopic approaches to try to minimize sources of random variability in the simulation. \\

In our simulation, 100 units of a covariate $z$ are generated. The covariate can take values in $\left\{-1,1\right\}$ and is generated such that $\mathbb{P}(z_i=1)=0.5$ and  $\mathbb{P}(z_i=-1)=0.5$ for all $i$. We assume the true model for the response is $y_i \sim $ Bernoulli($\pi_i$) with $\mbox{logit}(\pi_i)= z_i + t_i$, and generate responses according to this model. Our simulation is constructed as follows: 

\begin{enumerate}[I]
\item { \begin{enumerate}
\item 100 subjects are assumed and their covariates are generated. 
\item 100 deviates from a Unif$(0,1)$ distribution are generated for the response.
\item An initial design with 10 units is constructed using an exchange algorithm with $D_A$ optimality as the objective function.
\item Seven sequential designs using the covariates, random deviates for the responses, and initial design in part (a) are constructed using: 
\begin{itemize} 
\item A myopic $D_A$-optimal design.
\item A nonmyopic $D_A$-optimal design with horizon $N=1$, with the correct covariate distribution assumed. 
\item A nonmyopic $D_A$-optimal design with horizon $N=1$, with the empirical covariate distribution assumed. 
\item A nonmyopic $D_A$-optimal design with horizon $N=2$, with the correct covariate distribution assumed. 
\item A nonmyopic $D_A$-optimal design with horizon $N=2$, with the empirical covariate distribution assumed. 
\item A nonmyopic $D_A$-optimal design with horizon $N=3$, with the correct covariate distribution assumed. 
\item A nonmyopic $D_A$-optimal design with horizon $N=3$, with the empirical covariate distribution assumed. 
\end{itemize} 
\item Designs are evaluated using the performance measure $\Psi_{D_A}$, given by Equation \eqref{D_A_logit}, at each sample size between 10 and 100, inclusive. The true values of the parameters are used to calculate $\Psi_{D_A}$. 
\end{enumerate} }
\item (a)-(e) above is above 20 times to obtain a distribution of the performance measure for each sample size.
\end{enumerate}

In addition to comparing the estimates of $\bm{\beta}$ and the values of $\Psi_{D_A}$ for the myopic and non-myopic designs, we also consider the efficiency of the nonmyopic design relative to the myopic design. We define the $D_A$-efficiencies of a design $\bm{X}_i$ relative to another design $\bm{X}_i^*$ with parameter values $\bm{\beta}$ in the logistic model case as

\begin{equation}
 \label{DAeff_logit}
 \mathrm{Eff}_{D_A}= \left\{ \frac{\Psi_{D_A} \left( \bm{X_i^*, \bm{\beta} }\right) }{ \Psi_{D_A} \left( \bm{X}_i, \bm{\beta} \right) } \right\}^ {1/m}, 
 \end{equation}

where $m$ is the number of non-zero rows in the matrix $\bm{A}$.  \\

\begin{figure}[H]
		\centering
		\includegraphics[scale=0.7]{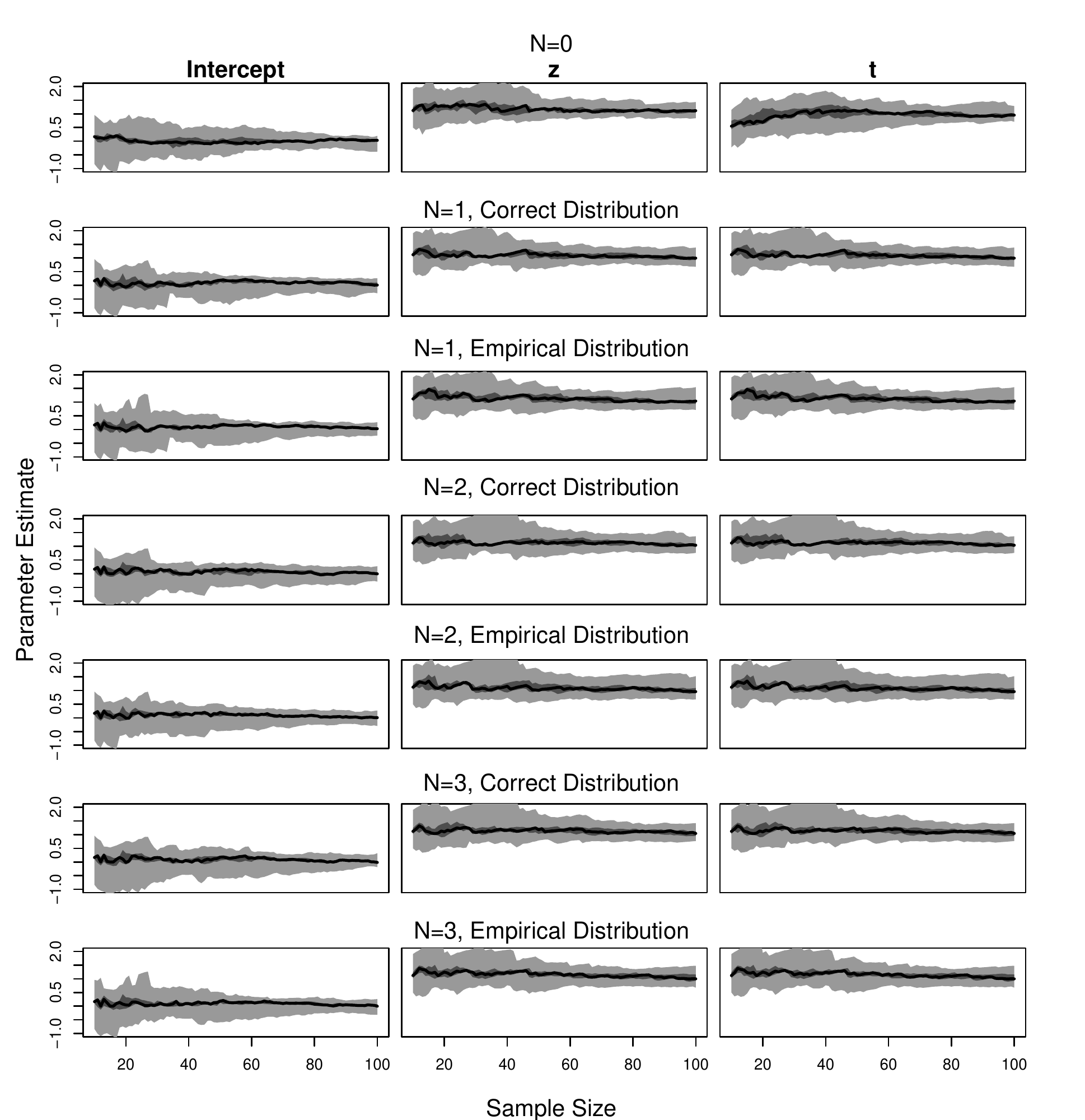}
		\caption[Comparisons of distributions of $\hat{\bm{\beta}}_i$ for myopic and nonmyopic $D_A$-optimal designs for the logistic model with one covariate]{Distributions of  $\hat{\bm{\beta}}_i$ for designs for the logistic model for one covariate are plotted against sample size. We show the myopic approach ($N=0$), as well as the nonmyopic approach to constructing $D_A$-optimal designs with horizon $N=$ 1 and 3. For the nonmyopic approach, we consider both the case where the correct covariate distribution is known, and when it is unknown so the empirical covariate distribution is used. The black line indicates the median, the dark grey indicates the 40th to 60th percentile, and the light grey indicates the 10th to 90th percentile of the distribution.}
		\label{cov1DAbeta}
\end{figure}

Figure \ref{cov1DAbeta} displays the distributions of  $\hat{\bm{\beta}}_i$ at each sample size between 11 and 100. The estimates appear to be centered around their true value, $\bm{\beta}=\left( 0, 1, 1\right) ^T$, and the plots appear to be very similar across the seven methods. 


\begin{figure}[H]
		\centering
		\includegraphics[scale=0.7]{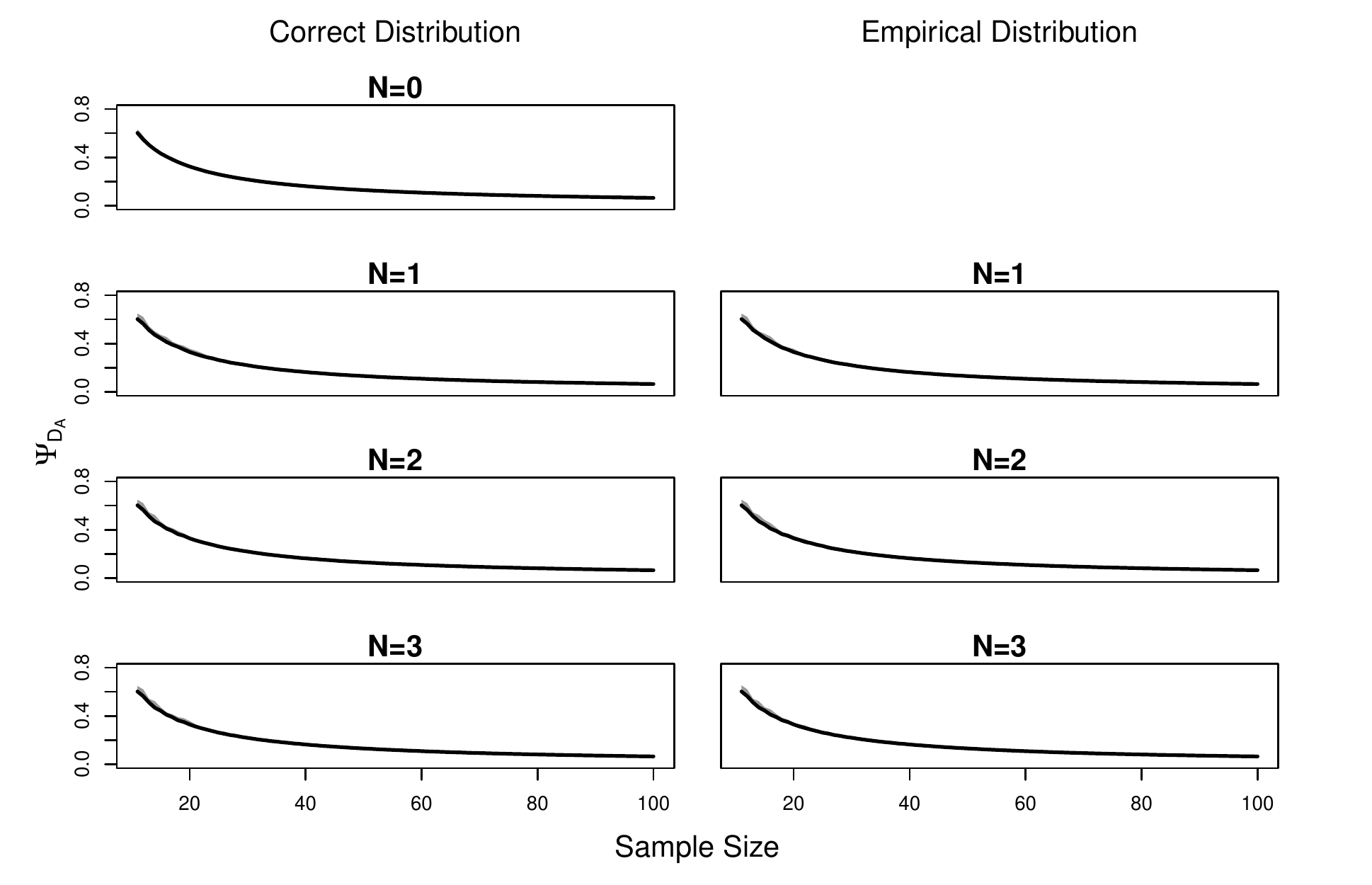}
		\caption[Comparisons of distributions of $\Psi_{D_A}$ for myopic and nonmyopic $D_A$-optimal designs for the logistic model with one covariate]{Distributions of $\Psi_{D_A}$ for designs for the logistic model for one covariate are plotted against sample size. We show the myopic approach ($N=0$), as well as the nonmyopic approach to constructing $D_A$-optimal designs with horizon $N=$ 1 and 3. For the nonmyopic approach, we consider both the case where the correct covariate distribution is known, and when it is unknown so the empirical covariate distribution is used. }
		\label{DAnonmy}
\end{figure}

\begin{figure}[H]
		\centering
		\includegraphics[scale=0.7]{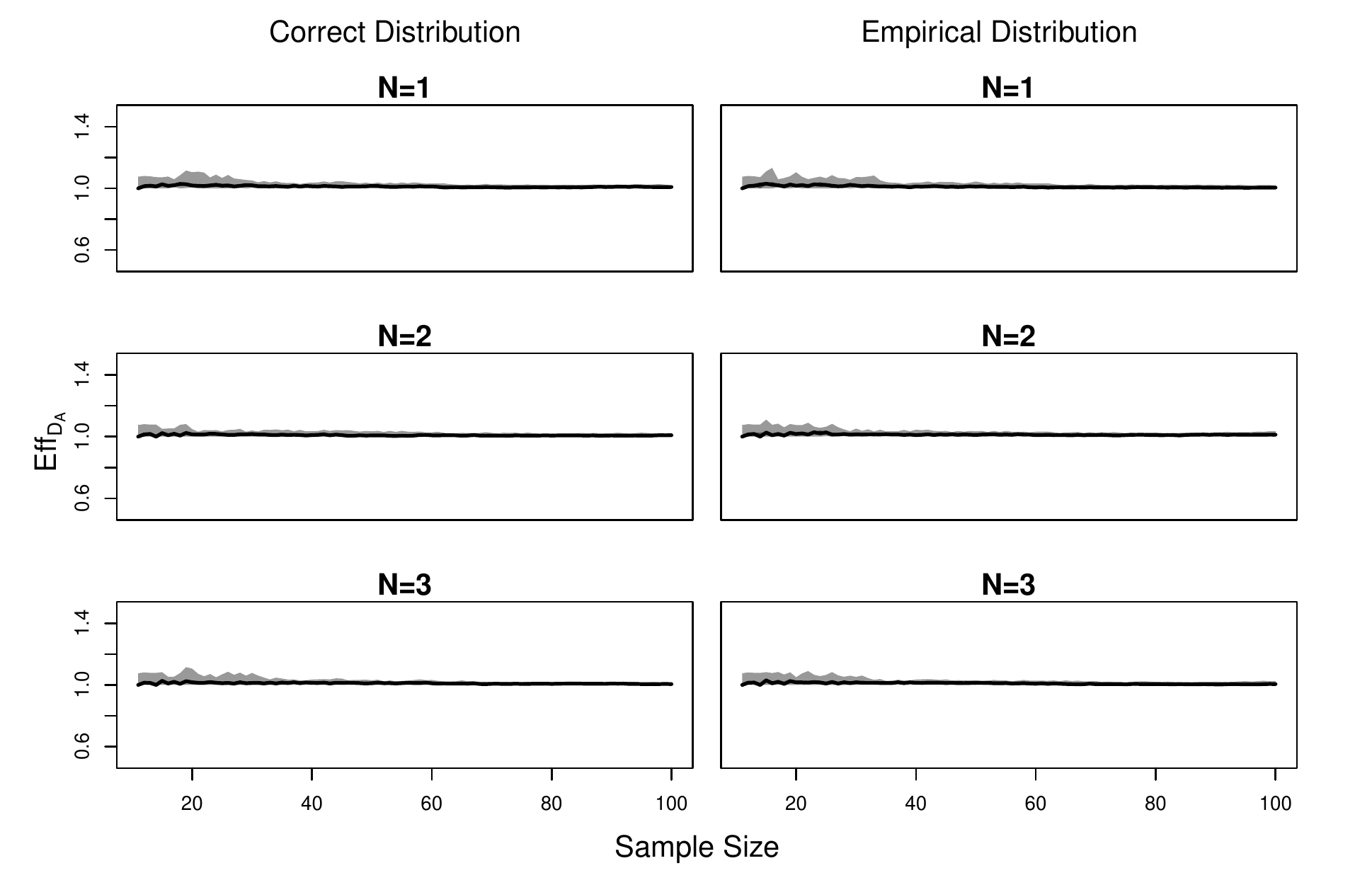}
		\caption[Comparisons of distributions of the relative efficiencies of the nonmyopic $D_A$-optimal  designs against the myopic $D_A$-optimal designs for the logistic model with one covariate]{Distributions of the relative efficiencies of the nonmyopic $D_A$-optimal designs against the myopic $D_A$-optimal designs for the logistic model for one covariate are plotted against sample size. We consider the efficiencies of the non-myopic approach with horizons 1 and 3, with the correct and empirical distributions, against the myopic approach as the baseline. }
		\label{DAnonmyeff}
\end{figure}

In Figure \ref{DAnonmy}, we plot the distribution of $\Psi_{D_A}$ for each sample size between 11 and 100. We observe that the value of this objective function decreases as sample size increases, as expected. We note that the plots look extremely similar across the seven methods. There is is no noticeable difference between having horizon equal to one or three. In Figure \ref{DAnonmyeff}, we plot the relative efficiencies of the nonmyopic designs against the myopic design, which confirms that there is no observable difference across the methods in $\Psi_{D_A}$; Table \ref{sim1tab} shows the efficiencies at the end of the experiment, and we see that the lower bound of the $10\%-90\%$ intervals are above 1. We observe that the myopic approach is slightly more efficient for when sample size is below 30. 

\begin{table}[H]
\caption{Distribution of the efficiencies of the non-myopic approaches relative to the myopic approach at the end of the experiment (n=100)}
\begin{tabular}{|l|l|l|l|l|}
\hline
\textbf{\begin{tabular}[c]{@{}l@{}}Efficiencies \\ when $n=100$\end{tabular}} & \textbf{median} & \textbf{40-60\% interval} & \multicolumn{2}{l|}{\textbf{10-90\% interval}} \\ \hline
$N=1$, correct dist & 1.008978 & (1.005972, 1.010633) & \multicolumn{2}{l|}{(1.000475, 1.019424)} \\ \hline
$N=1$, empirical dist & 1.005368 & (1.004617, 1.005720) & \multicolumn{2}{l|}{(1.000475, 1.019424)} \\ \hline
$N=2$, correct dist & 1.009279 & (1.005669, 1.012720) & \multicolumn{2}{l|}{(1.000342, 1.020574)} \\ \hline
$N=2$, empirical dist & 1.012895 & (1.009806, 1.013174) & \multicolumn{2}{l|}{(1.000475, 1.035371)} \\ \hline
$N=3$, correct dist & 1.005409 & (1.005232, 1.007396) & \multicolumn{2}{l|}{(1.000073, 1.016454)} \\ \hline
$N=3$, empirical dist & 1.005409 & (1.002690, 1.009182) & \multicolumn{2}{l|}{(1.000477, 1.024159)} \\ \hline
\end{tabular}
\label{sim1tab}
\end{table}

We observe in this simulation that there appears to be no benefit to the nonmyopic approach in this setting where we have one binary treatment and one binary covariate, and the covariate is generated such that $\mathbb{P}\left( z_i \right) = 0.5$ for all $i$. We call this a static covariate, since its distribution does not change with $i$. In Section \ref{pseudo_simluations}, we consider a dynamic covariate, where the distribution of the covariate changes over time.

\section{Pseudo-nonmyopic approach}

\label{pseudononmyopic}

One main limitation of the nonmyopic approach is that computing the nested expectations and minimizations over unknown quantities, such as in Equation \eqref{logisfuture}, requires recursive formulae which are computationally expensive. The number of calculations increases exponentially with each additional future subject in the horizon and, as a result, our simulations considered examples with horizon no more than three. We now explore a \textit{pseudo-nonmyopic} approach which involves evaluating a related objective function with a similar aim without the use of recursion. The computational burden is reduced as nested expectations and minimizations are not necessary but we are still able to incorporate information about future possible decisions. We describe this novel approach for the logistic model case (it can easily be described for the linear model case) and provide a simulation to show how it compares to the myopic approach.\\

In the pseudo-nonmyopic approach, in order to make a decision about the treatment of the $i$th subject, we generate $M$ possible \textit{trajectories} of covariate values for subject $i+1$ until subject $n$. We assume, as for the non-myopic approach, that we have a distribution $f_{\bm{z}}$ for the covariate $\bm{z}$. This may be the true distribution in the population (if it is known), or an empirical approximation based on the subjects in the trial up until the $i$th subject. The covariate distribution may depend on time, in which case we refer to it as a dynamic covariate. For each of the $M$ trajectories, we construct a \textit{pseudo-design} in which we have the $i$ subjects and $(n-i-1)$ subjects in the trajectory, and treatments allocated using an approach that we describe below. We look at the average losses of the $M$ pseudo-designs where we assign $t_i=1$, and compare it to the average loss of the $M$ pseudo-designs when $t_i=-1$; we select $t_i$ according to a probability that is weighted by these average losses. \\

This approach takes averages over simulated values of the covariates for subjects $i+1$  up to $n$. Optimization based on Monte Carlo simulations of unknown quantities is typically conducted in a Bayesian setting for design of experiments \citep{Woods2017}, where values of the unknown parameters may be simulated from a prior distribution. See \citet{Gentle2002} for an overview of Monte Carlo methods and \citet{Ryan2003} for an application to Bayesian design of experiments. \\

In order to create a design using the pseudo-nonmyopic approach for the logistic model, just like in the sequential myopic and nonmyopic algorithms, we begin by constructing an initial design $\bm{X}_{n_0}$. This involves an exchange algorithm where we assume $\bm{\beta}=\bm{0}$ as an initial guess. We then generate responses $\bm{y}_{n_0}$, and fit the model to obtain the initial estimates $\hat{\bm{\beta}}_{n_0}$. \\

Then, to select a treatment for subject $i$, for $i \in \left\{n_0+1, ... , n \right\}$, we observe $\bm{z}_i$. Based on the assumed covariate distribution $f_{\bm{z}}$,  we generate $M$ possible trajectories for the covariates, $\bm{z}_{(i+1):n}^1, \bm{z}_{(i+1):n}^2, ..., \bm{z}_{(i+1):n}^m$, where 

\begin{equation}
 \bm{z}_{(i+1):n} ^m = \begin{pmatrix} 
 z_{i+1}^m,  z_{i+2}^m, ..., z_{n}^m
 \end{pmatrix}^T,
\end{equation}

for $m \in \left\{ 1, 2, ..., M \right\}$. We then allocate treatments sequentially along each trajectory.\\

 Given the first subject in the trajectory, $\bm{z}_{i+1}^m$, we choose the treatment $t_{i+1}^{*^m}$ which minimizes the objective function $ \Psi$ given $t_i$, and the treatments and covariates of previous subjects and estimates $\hat{\bm{\beta}}_{i-1}$ based on the responses of the previous subjects, $\bm{y}_{i-1}$:

\begin{equation}
\label{logit_pseu}
t_{i+1}^{*^m} \left(  \bm{z}_{i+1}^m, t_{i} \mid \bm{z}_{i},  \bm{t}_{i-1}, \bm{y}_{i-1} \right)
=   \argmin _{t_{i+1}} 
 \Psi \left(  t_{i+1} \mid  \bm{z}_{i}, \bm{z}_{i+1}^m,  \bm{t}_{i-1}, t_i ,  \bm{y}_{i-1}\right). 
\end{equation}

To allocate a treatment for the next subject in the trajectory with covariate values $\bm{z}_{i+2}^m$, we then assume that $t_{i+1}^{*^m}$ has been allocated to subject $\bm{z}_{i+1}^m$ and choose the treatment $t_{i+2}^{*^m}$ which minimizes the objective function. We make the assumption that the future decisions are independent of the future responses. This means that we assume the same estimate for $\bm{\beta}$ as in the Equation \eqref{logit_pseu} and do not update it. We continue in this way until all subjects in the trajectory have been allocated a treatment:

For each $j$ in $\left\{ i+2, i+4, ..., n \right\}$, we define: 
\begin{align}
&t_{j}^{*^m} \left(  \bm{z}_{j}^m, t_{j-1}^* \mid \bm{z}_{i}, \bm{z}_{(i+1):(j-1)}^m, \bm{t}_{i-1},t_i, t_{(i+1):(j-2)}^*,  \bm{y}_{i-1}\right) \nonumber \\ 
&=   \argmin _{t_{j}} 
 \Psi \left(  t_{j} \mid  \bm{z}_{i}, \bm{z}_{(i+1):(j)}^m, \bm{t}_{i-1},t_i, \bm{t}_{(i+1):(j-1)}^* ,  \bm{y}_{i-1}  \right). 
\end{align}

For the $m$th trajectory, we obtain a pseudo-design with $n$ subjects where the $i$th treatment is 1, as well as a pseudo-design where the $i$th subject receives treatment $-1$. We denote the objective function of the two designs as follows: 
\begin{equation}
\Psi \left(  t_{n} \mid  \bm{z}_{i}, \bm{z}_{(i+1):n}^m, \bm{t}_{i-1}, t_i =1, \bm{t}_{(i+1):(n-1)}^{*^m},  \bm{y}_{i-1} \right),
\end{equation}
\begin{equation}
\Psi \left(  t_{n} \mid  \bm{z}_{i}, \bm{z}_{(i+1):n}^m, \bm{t}_{i-1}, t_i =-1, \bm{t}_{(i+1):(n-1)}^{*^m} ,  \bm{y}_{i-1}\right).
\end{equation}

We define the average objective function for $i=n_0+1, ..., n-1$ across the $M$ designs, assuming, firstly, that $t_i=1$, and secondly, that $t_i=-1$, as: 

\begin{equation}
\overline{\Psi}(t_i=1) = \frac{1}{M} \displaystyle \sum_{m=1}^M   \Psi \left(  t_{n} \mid  \bm{z}_{i}, \bm{z}_{(i+1):n}^m, \bm{t}_{i-1}, t_i =1, \bm{t}_{(i+1):(n-1)}^{*^m} ,  \bm{y}_{i-1}\right),
\end{equation}
\begin{equation}
\overline{\Psi}(t_i=-1) = \frac{1}{M} \displaystyle \sum_{m=1}^M \Psi \left(  t_{n} \mid  \bm{z}_{i}, \bm{z}_{(i+1):n}^m, \bm{t}_{i-1}, t_i =-1, \bm{t}_{(i+1):(n-1)}^{*^m} ,  \bm{y}_{i-1}\right).
\end{equation}

For $i=n$, we do not generate any future covariates so we have:

\begin{equation}
\overline{\Psi}(t_i=t)  = \Psi \left(  t_{n}=t \mid  \bm{z}_{n}, \bm{t}_{n-1} ,  \bm{y}_{n-1} \right),
\end{equation}

for $t \in \left\{-1, 1\right\}$. \\

We sample $t_i$ from the set $\left\{-1, 1\right\}$ where the probability of selecting $1$ is given by 
 \begin{equation}
\frac{ \overline{\Psi}(t_i=1)^{-1} }{\overline{\Psi}(t_i=1)^{-1} + \overline{\Psi}(t_i=-1)^{-1}}.
 \end{equation}
We then observe the response $y_i$ and refit the model to obtain $\bm{\hat{\beta}}_i$. \\

\subsection{Simulations}
\label{pseudo_simluations}

Similarly to Section \ref{nonmy_simulations}, we need to make sure that sources of variability are controlled as much as possible so that differences between the results for the myopic and pseudo-nonmyopic approaches are likely to be attributable to the differences in the treatment allocation approach. We make sure that simulations have the same initial design; the initial design is constructed with an exchange algorithm to allocate treatments to 10 units, under the assumption that $\bm{\beta}$ is a vector of zeros. We fit the models using the \texttt{R} function \texttt{bayesglm} in the \texttt{arm} package \citep{Rarm}, with Cauchy prior distribution with center zero and scale given by 2.5 for both the treatment and covariate parameters. We generate deviates $u_i$ as in Equation \eqref{ueq} in order to generate responses $y_i$, so that we can ensure that the data generating mechanism is the same across simulations comparing the myopic and pseudo-nonmyopic designs. \\

In this example, we have one binary covariate ${z}$. It is dynamic with a distribution given by $\mathbb{P}(z_i=1) =0.01i$. The model is given by ${y}_i \sim$ Bernoulli$(\pi_i)$ where

\begin{equation}
\mbox{logit}\left(\frac{\pi_i}{1-\pi_i} \right) =   \ {z}_i +  {t}_i.
\end{equation}

The structure of the simulation is as follows: 

\begin{enumerate}[I]
\item { \begin{enumerate}
\item 100 subjects are assumed and their covariates are generated from a specified distribution. 
\item 100 deviates from a Unif$(0,1)$ distribution are generated for the response.
\item An initial design with 10 units is constructed using an exchange algorithm with $D_A$-optimality as the objective function.
\item The three following sequential designs are constructed using the covariates, random deviates for the responses, and initial design in part (a):  
\begin{itemize} 
\item A myopic $D_A$-optimal design.
\item A pseudo-nonmyopic $D_A$-optimal design with $M=10$, and the correct covariate distribution assumed. 
\item A pseudo-nonmyopic $D_A$-optimal design with $M=100$, and the correct covariate distribution assumed. 
\end{itemize} 
\item Designs are evaluated using the performance measure $\Psi_{D_A}$ at each sample size between 10 and 100, inclusive. The true values of the parameters are used to calculate $\Psi_{D_A}$. 
\end{enumerate} }
\item (a)-(e) above is repeated 20 times to obtain a distribution of the performance measure for each sample size.
\end{enumerate}

\begin{figure}[H]
		\centering
		\includegraphics[scale=0.7]{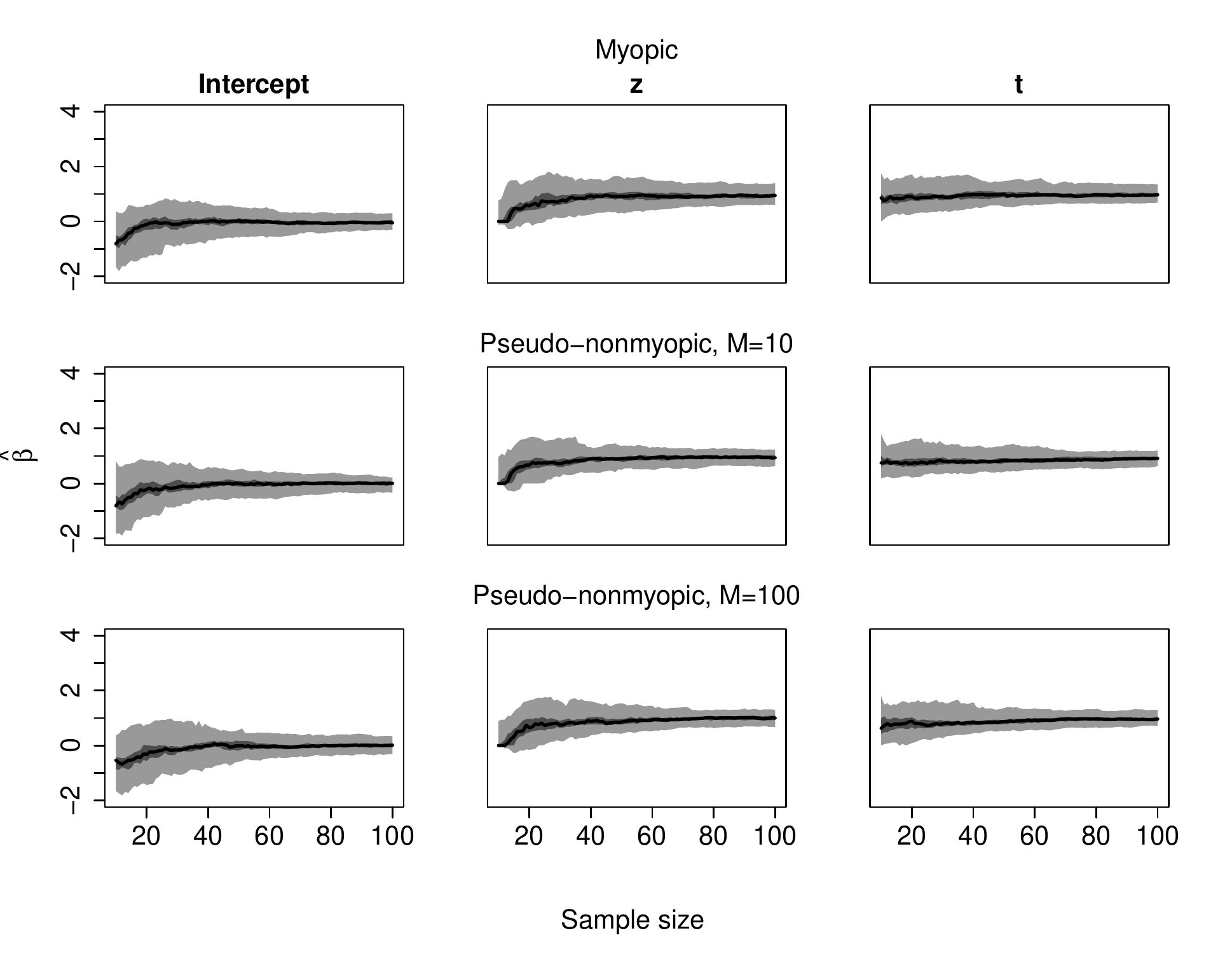}
		\caption[]{Parameter estimates given by the myopic approach, pseudo-nonmyopic approach with $M=10$ and pseudo-nonmyopic approach with $M=100$ for a logistic model with one dynamic
covariate. The black line indicates the median, the dark grey indicates the 40th to 60th percentile, and the light grey indicates the 10th to 90th percentile of the distribution}
		\label{011pseudo_beta}
\end{figure}

In Figure \ref{011pseudo_beta}, we see the estimates of $\bm{\beta}$ for the myopic approach, the pseudo-nonmyopic approach with $M=10$ and with $M=100.$ We observe that the plots looks very similar across the three methods. The variability of the estimates reduces with sample size for the intercept and the coefficient of treatment. The median of the distributions converge to their true value after a sample size of approximately 40. 

\begin{figure}[H]
		\centering
		\includegraphics[scale=0.7]{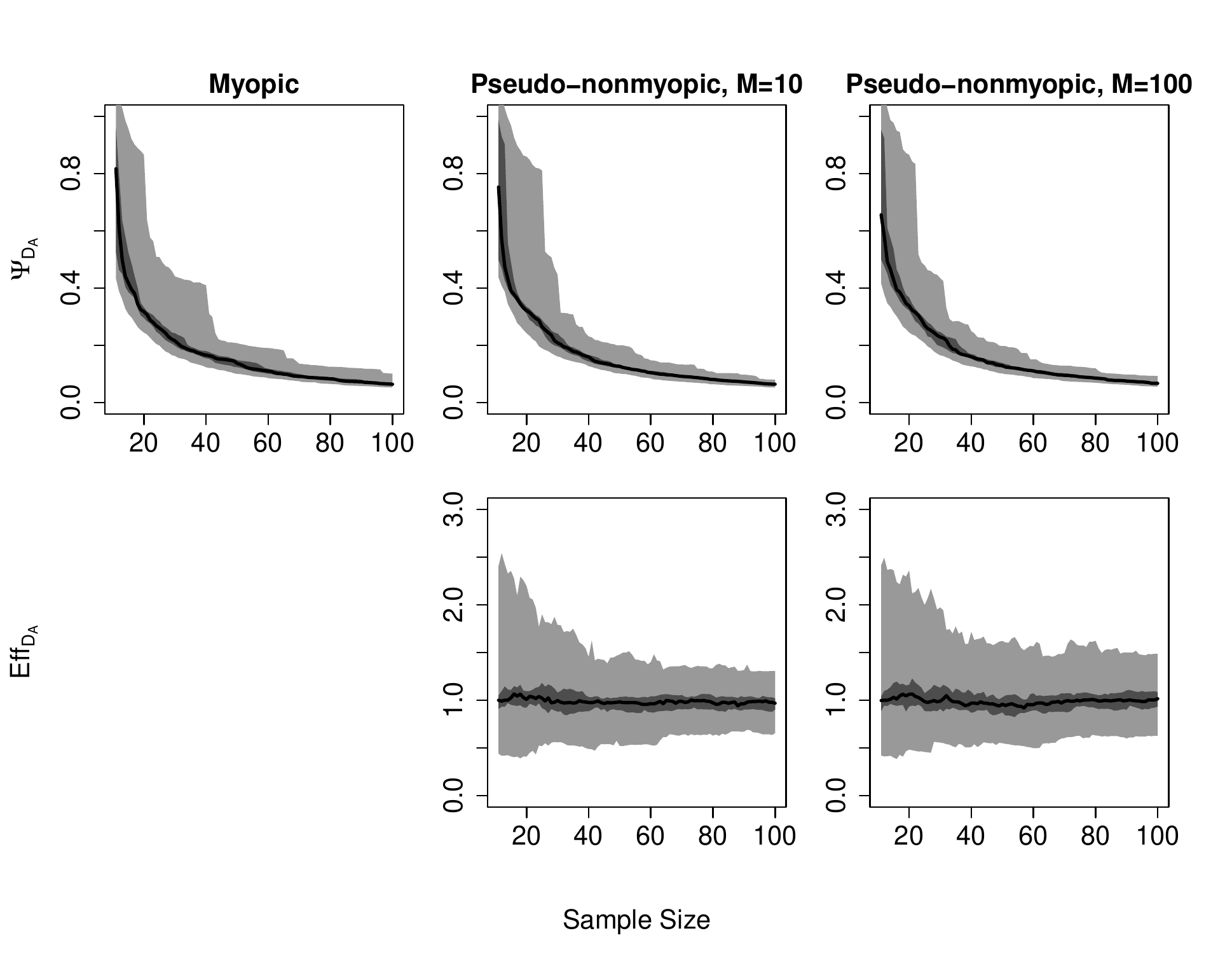}
		\caption[]{Top row: $D$-optimality against sample size for designs for a logistic model with one dynamic
covariate. Bottom row: relative $D$-optimality against sample size for designs for a logistic model with one
dynamic covariate. Values below 1 indicate that the pseudononmyopic approach is more beneficial than
the myopic approach.}
		\label{011pseudo_relopt}
\end{figure}

In Figure \ref{011pseudo_relopt},  the top row displays the values of $\Psi_{D_A}$ evaluated at each sample size. This appears to be similar across all methods with slightly higher variation observed for the pseudo-nonmyopic approach with $M=10$. In all three cases, the value of the objective function drops after a few initial subjects and stabilizes after around 30 subjects. The bottom row shows the relative $D_A$-efficiencies (see Equation \eqref{DAeff_logit}) of the pseudo-nonmyopic approaches, compared to the myopic approach. We see that, initially, they have equal efficiency, but then the myopic approach appears to be slightly more efficient. We note that the distributions of efficiencies are skewed; there appears to be a number of extreme points where the myopic approach is much more efficient than the pseudo-nonmyopic approach. This is partly due to the fact that the efficiency is bounded below by zero, but unbounded above. Table \ref{sim2tab} displays the efficiencies at the end of the experiment; the distributions are  centered around one and have greater uncertainty than in the efficiencies of nonmyopic approach in section Table \ref{sim1tab}. 

\begin{table}[H]
\caption{Distribution of the efficiencies of the pseudo-nonmyopic approaches relative to the myopic approach at the end of the experiment (n=100)}
\centering
\begin{tabular}{|l|l|l|l|l|}
\hline
\textbf{\begin{tabular}[c]{@{}l@{}}Efficiencies \\ when $n=100$ \end{tabular}} & \textbf{median} & \textbf{40-60\% interval} & \multicolumn{2}{l|}{\textbf{10-90\% interval}} \\ \hline
$M=10$ & 0.9690018 & (0.9014291, 1.0202391) & \multicolumn{2}{l|}{(0.6544362, 1.3059580)} \\ \hline
$M=100$ & 1.0157450 & (0.9340631, 1.0845578) & \multicolumn{2}{l|}{(0.6267983, 1.4871852)} \\ \hline
\end{tabular}
\label{sim2tab}
\end{table}

We found no evidence for the benefit of the pseudo-nonmyopic approach over the nonmyopic approach in this example. Further, we observed that the number of trajectories in the pseudo-nonmyopic approach, $M$, appears to have little effect on the parameter estimates or the values of the $D_A$-optimal objective functions.

\section{Discussion}
\label{discussion}

This paper extended  the sequential optimal design approach first proposed by \cite{Atkinson1982} for the logistic model case and for any optimality criterion. We then placed this approach in a nonmyopic framework. In our simulations, we observed no benefit to using the nonmyopic approach over the myopic approach. We then developed a novel methodology called the pseudo-nonmyopic approach  which is still able to take into account future possible subjects, but is less computationally expensive than the nonmyopic approach. Simulations showed that the pseudo-nonmyopic approach performs similarly to the myopic approach for the logistic model case with a binary treatment.

\subsection{Limitations} 
There are a number of limitations to our work in its ability to be directly applicable to clinical trials and other experiments involving human subjects. Firstly, we assume responses are measured immediately after treatments are given to subjects. This is not a realistic assumption so some method to allow for a delay between treatment allocation and response could be useful. One modification would be to allow for the method to be batch sequential; instead of allocating treatments to one subject at a time, a group of subjects may be given optimal treatments by using the exchange algorithm.  It is also possible to incorporate delay in adaptive designs. \citet{Hardwick2006} achieve this by assuming that subjects arrive according to a Poisson process. \\

Secondly, we do not consider toxicity in our work. We assume that the treatment which leads to a better response is the more desirable treatment, but it is possible that such a treatment has unsafe toxicity levels \citep{Rosenberger1999}. In our algorithms for treatment assignment, if the optimality criterion is equal for treatment $t_i=1$ and $t_i=-1$, we would assign the treatment at random. In clinical trials, this is less likely to happen as relative efficiency of the treatments need to be considered in conjunction with relative toxicity \citep{Simon1977}. In general, \cite{Rosenberger1999} recommended that adaptive designs should be considered after previous experiments have been able to establish low toxicity of the treatments. \\

A further limitation of our work is that we arbitrarily assume in all of our simulations that we have 100 subjects in the trial. In clinical trials, there are stopping rules that determine when the trial should terminate \citep{Stallard2001}. See \citet{Whitehead1993} for a frequentist perspective and \citet{Berry1989} and \citet{Freedman1989} for a Bayesian perspective on stopping rules in interim analysis. Including this element into our designs would mean that our methodology is more generally applicable to clinical trials. Further, we may be able to make statements about relative numbers of subjects and costs required in order to detect a significant difference in treatment effect for each method.

\subsection{Future Work}
The non-myopic and pseudo-nonmyopic algorithms consider only the case where the response and treatments are binary. Natural extensions include allowing for more complex treatment structures, such as factorial designs, or allowing for a continuous response. Computing the expected objective function for a continuous response would require Monte-Carlo simulations. Extending our framework for the non-myopic approach to allow for a more general response will require greater computational efficiency in our algorithms. This is also true of the pseudo-nonmyopic approach. \\

In the optimality criteria that we have considered, the response of the subjects are included in order to update parameter estimates (optimal design methods for the logistic model case, weighted $L$-optimal design). The response has not been used in order to inform treatment allocation based on the efficacy of the treatment. Covariate adjusted response-adaptive designs based on efficiency and ethics (CARAEE) aim to optimize a utility function which takes into account the number of subjects who receive the more effective treatment. We did some preliminary work on CARAEE designs. Here, our optimality criterion is a function which has a component for efficiency and a component for ethics, as well as a tuning parameter which allows the practitioner to decide which aim is more important. The CARA (covariate adjusted response adaptive) design and RAR (response adaptive randomization) design are special cases of the CARAEE design.\\

\section*{Appendix A}
\label{appendix_code}
An $\texttt{R}$ package $\texttt{biasedcoin}$ is included in the Supplementary materials. The following commands implement the designs for logistic regression discussed in this paper: \\

$\texttt{logit.coord}$: non-sequential optimal design (coordinate exchange algorithm).\\
$\texttt{logit.des}$: myopic sequential optimal design. \\
$\texttt{logit.nonmy}$: nonmyopic sequential optimal design.\\
$\texttt{simfuture.logis}$: pseudo-nonmyopic sequential optimal design. \\

The \texttt{R} function \texttt{bayesglm} in the \texttt{arm} package is used to fit the logistic regression model using the Cauchy prior to avoid problems in estimation due to separation.

\section*{Appendix B}
  \label{appendix_algorithm}

\begin{algorithm} [H]                     
\caption[Myopic sequential optimal design for binary response]{Myopic sequential optimal design for binary response:  this algorithm returns a design matrix, constructed sequentially. The inputs are the covariate values, $\bm{z}_{n}$, and the number of subjects in initial design, $n_0$. }
\label{Algorithm1}
\begin{algorithmic}[1]
	\Function{SeqOptL}{$\bm{z}_n, n_0 $} 
	\newline
		\State \textbf{Initialization} 
		\State Construct initial design  $\bm{X}_{n_0}$ using the exchange algorithm for the first $n_0$ subjects assuming $\bm{\beta}=\bm{0}$.
		\State Observe responses $\bm{y}_{n_0}=\begin{pmatrix} y_1, y_2, ..., y_{n_0} \end{pmatrix} ^T.$
		\State Fit the model assuming that the responses are distributed according to Equation \eqref{dist} with linear predictor given by Equation \eqref{predictor} to obtain the MLE $\hat{\bm{\beta}}_{n_0}$. 
		\newline 
		\For{$i$ in $n_0 +1$ to $n$ }
		 \State Observe  $z_{i,1}, ..., z_{i, k}$ 
		 \State Calculate  $\Psi (t_{i} \mid \bm{z}_{i},  \bm{t}_{i-1}, \bm{y}_{i-1})$ for each treatment.
		 \State Sample treatment for subject $i$ where probability of treatment $1$ is given by Equation \eqref{logit_tmt1}.
		 \State Observe response $y_i$.
		 \State Refit model the model with response given by $\bm{y}_{i}$ and updated design matrix $ \bm{X}_{i}$ and update the parameter estimates $\hat{\bm{\beta}}_i$.
		\EndFor
		\newline 
	\State \Return $\bm{X}$	\newline
	\EndFunction
\end{algorithmic}
\end{algorithm}


\newpage
\bibliographystyle{rss}
\bibliography{library}

\end{document}